\documentclass[lettersize,journal]{IEEEtran}

\usepackage[T1]{fontenc}
\usepackage{amsmath,amsfonts,amssymb,amsthm}
\usepackage{bm}

\usepackage{graphicx}
\usepackage{psfrag}
\usepackage{stfloats}
\usepackage[caption=false,font=normalsize,labelfont=sf,textfont=sf]{subfig}

\usepackage{algorithm}
\usepackage{algorithmic}

\usepackage{array}
\usepackage{booktabs}

\usepackage{cite}

\usepackage{textcomp}
\usepackage{url}
\usepackage{verbatim}

\begin{document}
\title{Noncoherent Detection of Constant-Envelope Signals for Mobile Edge Applications -- Optimum Detectors and Intelligent Decision Rule}

\author{Mu Jia, \IEEEmembership{Graduate Student Member,~IEEE}, Junting Chen, \IEEEmembership{Member,~IEEE}, Ying-Chang Liang, \IEEEmembership{Fellow,~IEEE}, and Pooi-Yuen Kam, \IEEEmembership{Life Fellow,~IEEE}
	
	\thanks{
		Mu Jia, Junting Chen and Pooi-Yuen Kam are with Guangdong Provincial Key Laboratory of Future Networks of Intelligence, School of Science and Engineering, Shenzhen Future Network of Intelligence Institute (FNii-Shenzhen), The Chinese University of Hong Kong, Shenzhen, Guangdong, P.R. China (e-mail: mujia1@link.cuhk.edu.cn; juntingc@cuhk.edu.cn; pykam@cuhk.edu.cn).
		
		Ying-Chang Liang is with the Center for Intelligent Networking and Communications, University of Electronic Science and Technology of China, Chengdu 611731, China (e-mail: liangyc@ieee.org).
	}
}

\maketitle

\begin{abstract}
	Constant-envelope signals are widely used in mobile edge applications and wireless communication systems for their hardware-friendly design, energy efficiency, and reliability. However, reliable detection with simple, power-efficient receivers remains challenging. Coherent methods offer superior performance but require complex synchronization, increasing complexity and power use. Noncoherent detection is simpler, avoiding synchronization, but traditional approaches rely on in-phase and quadrature-phase (IQ) demodulators for signal magnitudes and assume energy detectors without theoretical justification.
	This paper proposes a framework for optimal detection using a bandpass-filter envelope-detector (BFED) with Bayes criterion and generalized likelihood ratio test (GLRT) under unknown amplitudes. Using modified Bessel function approximations, we show the optimal detector shifts based on SNR: in the low-SNR regime, we rigorously prove for the first time that the well-known energy detector (ED) is the Bayesian-optimal solution, thus providing a firm theoretical foundation for its widespread use; in high-SNR regimes, a novel amplitude detector (AD) compares estimated amplitude to noise deviation, leading to a simple yet optimal detection strategy.
	For unknown SNR, a reliability-based intelligent decision (RID) rule adaptively selects detectors, leveraging their strengths across SNR ranges. Simulations confirm energy and amplitude detectors minimize errors in their domains, with RID providing robust gains. The proposed framework provides a rigorous theoretical foundation and enables low-complexity implementations for resource-constrained, interference-limited mobile edge applications, including wireless sensor networks (WSNs) and Internet of Things (IoT) systems.
\end{abstract}


\section{Introduction}
\IEEEPARstart{S}{ignal} detection plays a crucial role in modern communication systems and mobile edge applications, ensuring reliable and efficient data transmission. The ability to detect signals accurately amidst noise and interference is fundamental, particularly as the mobile communication systems continue to evolve in complexity and capability \cite{WeiQuWan:J23,PhaDinKim:J23,DaiWanYua:J15,KumDas:J20,CheZhaFen:J22,YeLiJua:J18,YuaLonWan:J23,WanGaoFan:J16,HuaLuoZha:J22,KhaAliHoyFle:J20,YanHan:J15,HeWenJinLi:J20}.

In the realm of signal detection, much attention has been given to signals with arbitrary envelopes, random distributions, and various other characteristics \cite{Che:J10,CheFen:J24,liu2020deep}. Here, we focus on the detection of constant-envelope signals due to their practical applications and importance in communications systems and mobile edge applications \cite{MalRanMur:J21,MalMur:J24,MonHanNgu:J22,digham2003energy,ZhaZhaLim:J18}. Constant-envelope signals, which maintain a consistent amplitude irrespective of the information they convey, are vital across several domains due to their advantages in hardware friendliness, energy efficiency, and communication reliability.
A foundational application is spectrum sensing for cognitive radio, where simple and power-efficient methods to detect primary user (PU) signals are necessary. Many PU systems, such as frequency modulated (FM) radio and certain digital mobile radio standards, utilize constant-envelope modulation. The stable amplitude of these signals greatly simplifies the design of receivers, directly leveraging their inherent energy efficiency to enable robust and low-complexity dynamic spectrum access \cite{TahAmiAbb:J25,YenChe:J20,wuXuZho:J23,LuoHuaChe:J24}.
In addition, in IoT applications, edge devices require low-power, simple receivers for signal detection. This requirement motivates the use of constant-envelope signals compatible with energy-efficient amplification, making them particularly suitable for activation signals \cite{ChougraniIoT,yan2018analysis,Zhang2020}.
In WSNs, constant-envelope signals are widely used to enable efficient device activation and localization. Their properties enhance range and sensing capabilities while conserving power and extending system life time \cite{shenyuanWSN,YinHas:J19,Jia}.

Typically, coherent detection methods generally offer superior performance in signal detection, as they can extract both amplitude and phase information from the received signal. In systems using IQ demodulators, the incoming modulated signal is split into its in-phase and quadrature components by mixing with two local oscillator signals that are 90 degrees out of phase with each other. This allows for precise amplitude and phase recovery, enabling efficient signal processing \cite{Urk:J67,digham2003energy,Che:J10,AlAlPar:J20,CheFen:J24,liu2020deep,Kam1994}. 
However, this matching process introduces significant complexity and resource demands in hardware design, particularly due to the need for precise frequency and phase matching to ensure proper separation and reconstruction of the IQ components of the signal. 

With the increasing demand for simpler, resource-efficient receivers, particularly for the noncoherent detection of constant-envelope signals where devices often only need to detect the presence of signals rather than demodulate all information \cite{MalRanMur:J21,HuaLanZha:J20,nondetambc}, alternative detection structures are sought. The BFED presents a compelling simplified receiver design \cite{van2004detection}. In this paper, we demonstrate that BFED can effectively detect constant-envelope signals by matching them with a known reference to maximize energy, thereby operating across a wide frequency band and eliminating the need for precise frequency matching. We also show that with the BFED, amplitudes can be estimated using only magnitudes with a simpler maximum likelihood estimator (MLE), thus significantly reducing the complexity of the receiver structure compared to traditional methods \cite{wu2021maximum}. These results illustrate how BFED provides a practical and efficient solution for constant-envelope signal detection, particularly in resource-constrained environments.

While BFED offers a practical solution for constant-envelope signal detection in resource-constrained environments, it raises an important theoretical question: is the energy detector truly the optimal noncoherent detection statistic, or could better decision metrics be derived? Traditional approaches often heuristically assume that the optimal noncoherent detector reduces to a simple energy detector, based on the structure of IQ demodulators, without a rigorous formal proof \cite{Urk:J67,digham2003energy,CheFen:J24,nondetambc}. This creates a critical gap in the theory, as it remains unclear whether more effective detection strategies can be formulated from first principles.
In this work, we revisit the noncoherent detection framework, beginning directly from the envelope-detected output—the sequence of noisy received signal magnitudes—while discarding any phase information. By carefully re-deriving the optimal detector from these magnitude-only observations, we not only provide the first rigorous justification for the energy detector’s optimality in the low SNR regime, but also uncover a novel amplitude-based detector structure that emerges at high SNR. This rigorous treatment not only resolves longstanding questions in noncoherent detection theory but also lays the groundwork for more efficient and insightful receiver designs tailored to constant-envelope signal communication.

Motivated by these considerations, we draw upon established statistical tests to systematically determine the optimum decision rules. Two statistical tests are generally utilized in this derivation: the Neyman-Pearson (NP) test and the Bayesian test \cite{van2004detection}. 
The NP test maximizes the detection probability for a specified false alarm rate, making it the most powerful test for balancing detection and false alarm trade-offs at each point along the Receiver Operating Characteristic (ROC) curve. However, it does not directly minimize the overall error probability, which can be a limitation in certain applications where both types of errors are equally important. If we have the prior knowledge about the probabilities of the hypotheses and the costs associated with different decision outcomes, Bayesian is a better approach, which can lead directly to the optimum decision threshold, and achieves the optimum probability of error. This approach allows us to design a detector that is more robust and efficient for a wide range of practical communication scenarios.

We assume that the noise power spectral density at the receiver is known, as it can be easily measured when no signal is present. However, in communication systems, the amplitudes often need to be estimated online due to the information or physical meaning they convey. For example, in Integrated Sensing and Communication (ISAC) systems, the estimation of signal amplitudes is related to target distance and reflection environments, making it essential to accurately estimate them \cite{WeiQuWan:J23,WeiLiuLi:J25,ZhoLenWan:J23,wei2023integrated}. Thus, the amplitude of constant-envelope signals needs to be estimated using MLE based on the magnitudes of the noisy received signals obtained from the previous simple receiver design, as investigated in our prior work \cite{wu2021maximum}. To address this, we employ the GLRT to derive the optimal detector, which allows for effective hypothesis testing even when signal parameters are not fully known. The unknown amplitudes are estimated using MLE, and a likelihood ratio test is then formed based on these estimates.

In this paper, we develop and analyse the procedures to achieve optimal detection of constant-envelope signals by leveraging the received signal magnitudes. We focus on a hardware-lightweight solution using the BFED to detect signal amplitudes, particularly in resource-constrained environments. Building on the constant-envelope signal model, we formulate the detection problem using Bayesian principles and the GLRT. The nonlinear nature of the GLRT leads us to approximate Bessel functions and separately address detection in the low and high SNR regimes, leading to distinct detection strategies for each.
For low SNR, we derive an energy detector, an approach commonly used in traditional signal detection. However, we make a significant contribution by for the first time providing a rigorous theoretical proof of the energy detector's optimality in this regime, underpinned by Bayesian test. This allows us to determine the precise detection threshold, ensuring maximum detection reliability while minimizing the error probability. In contrast, for high SNR, the optimum detector turns out to be a new detector that incorporates MLE of the signal amplitude. This detector compares the estimated amplitude to the noise standard deviation to perform detection, offering improved performance over traditional detectors in this regime.
Although each detector is optimal in its own operating region, neither is uniformly best across all SNRs, and in the transition region they may even yield conflicting decisions for the same observation block. To fully exploit their complementary strengths while retaining low implementation complexity, we further design a RID rule. The RID rule uses analytically characterized error probabilities of ED and AD to adaptively fuse their outputs, effectively selecting the more reliable decision without requiring explicit SNR knowledge.
Through a combination of analysis and extensive numerical simulations, we validate the effectiveness of these detectors across a range of noise conditions, demonstrating their robustness and the superior performance of the new high SNR detector and the RID detector. Our findings contribute to the field by offering a rigorous, unified approach to constant-envelope signal detection that bridges theory and practical implementation.

The paper is organized as follows. Section II describes the signal model and derives the optimal detection rule using the GLRT. Section III analyses the performance of ED and AD. Section IV introduces the proposed RID rule, detailing its decision cases and structural properties. Section V presents numerical results that compare ED, AD, and RID, and Section VI concludes the paper.

\section{Detection based on Received Signal Magnitude}

\subsection{Received Signal Model}
We consider the problem of detecting a constant envelope signal with circularly symmetric complex Gaussian (CSCG) noise.
The binary hypothesis testing problem of the received signal can be formulated as 
\begin{align}
	H_1:\quad& r(k) = A e^{j(\omega k + \theta)} + n(k),\quad k=0,1,\ldots,N-1 \\
	H_0:\quad& r(k) = n(k),\quad k=0,1,\ldots,N-1
	\label{hypothesis}
\end{align}
where $A$ is the amplitude of the signal, $N$ the number of samples, $\omega$ the frequency of the signal, $\theta$ the phase of the signal and $n(k)$ complex additive white Gaussian noise (AWGN) with zero mean and variance $2\sigma^2$, which is assumed as prior knowledge here\footnote{This is a standard and necessary assumption in detection theory. In practice, the noise power can be readily estimated by observing the channel during idle periods, and this estimation is also been investigated in our previous work \cite{wu2021maximum}. However, it is important to acknowledge that any error in this estimation will result in a suboptimal threshold, potentially increasing either the false alarm or misdetection rate. While a detailed robustness analysis of the detectors against noise estimation errors is a valuable topic for future investigation, this paper focuses on establishing the fundamental, optimal detection framework under the condition of known noise.}. $H_1$ assumes the presence of a signal, while $H_0$ represents the case where only noise is present.
The SNR level here is defined as $\rho = \frac{A^2}{2\sigma^2}$. 

For the simplified noncoherent receiver, we apply the BFED to the received signal model $r(k)$, which structure is a bandpass filter connected with an envelope detector. BFED can effectively extract the signal's envelope without the complexities introduced by frequency and phase variations. The end result of this process is $\left\{|r(k)|\right\}_{k=0}^{N-1}$, which is a simplified signal representation that emphasizes amplitude, easing subsequent processing steps. And we will derive the optimal detector from this representation.

From (\ref{hypothesis}), the distribution of the amplitude of received signal $|r(k)|$ under $H_1$ follows a Rician distribution with the parameters $A$ and $\sigma$. The joint probability density function of all the i.i.d. received samples $\left\{|r(k)|\right\}_{k=0}^{N-1}$ conditioned on the unknown amplitude $A$ can be derived as
\begin{align}
	\nonumber
&	P\left(\left\{|r(k)|\right\}_{k=0}^{N-1} \mid H_1,A\right)\\
&	=\prod_{k=0}^{N-1} \frac{|r(k)|}{\sigma^2} \exp\left({-\frac{|r(k)|^2+A^2}{2\sigma^2}}\right)I_0\left(\frac{|r(k)| A}{\sigma^2}\right)
\label{pdfh1}
\end{align}
where the $I_v(.)$ is the modified Bessel function of the first kind . Meanwhile, $|r(k)|$ under $H_0$ follows a Rayleigh distribution, and the joint PDF of samples is 
\begin{align}
	P\left(\left\{|r(k)|\right\}_{k=0}^{N-1} \mid H_0\right)=\prod_{k=0}^{N-1} \frac{|r(k)|}{\sigma^2} \exp\left({-\frac{|r(k)|^2}{2\sigma^2}}\right)
	\label{pdfh0}
\end{align}
while the corresponding log-likelihood functions are
\begin{align}
	\nonumber
	\label{H1LLH}
	H_1: \quad &\ln P\left(\left\{|r(k)|\right\}_{k=0}^{N-1} \mid H_1,A\right) \\
	\nonumber
	= &-N \ln (\sigma^2) + \sum_{k=0}^{N-1} \ln |r(k)| - \frac{N A^2}{2\sigma^2} \\
  &- \frac{1}{2\sigma^2} \sum_{k=0}^{N-1} |r(k)|^2 + \sum_{k=0}^{N-1} \ln I_0 \left( \frac{|r(k)| A}{\sigma^2} \right)\\
	\nonumber
	H_0: \quad	&\ln P\left(\left\{|r(k)|\right\}_{k=0}^{N-1} \mid H_0\right) \\
	= &-N \ln (\sigma^2) + \sum_{k=0}^{N-1} \ln |r(k)| - \frac{1}{2\sigma^2} \sum_{k=0}^{N-1} |r(k)|^2 .
\end{align}

\subsection{Generalized Likelihood Ratio Test}
According to the Bayesian test, we want to derive the optimum detector with the minimum probability of error
\begin{align}
	P_{\text{e}}=P(H_0|H_1)P(H_1)+P(H_1|H_0)P(H_0) .
\end{align}
To minimize the $P_{\text{e}}$, we apply the likelihood ratio test by comparing the likelihood ratio to a threshold. We assume an equal prior probability $P(H_1)=P(H_0)$. 

Due to that the amplitude of the received signal is unknown, the optimal detection performance can be achieved by the GLRT. Based on the likelihood calculated from  (\ref{pdfh1}) and (\ref{pdfh0}), the GLRT is given by \cite{van2004detection}
\begin{align}
\Lambda=\frac{ \mathop{max}\limits_A  P\left(\left\{|r(k)|\right\}_{k=0}^{N-1} \mid H_1,A\right)}{P\left(\left\{|r(k)|\right\}_{k=0}^{N-1} \mid H_0\right)} 
	\underset{H_0}{\overset{H_1}{\gtrless}} 1 .
\end{align}
By finding the MLE of the unknown amplitude $\hat{A}$, we maximizes the likelihood function of $H_1$, $P\left(\left\{|r(k)|\right\}_{k=0}^{N-1} \mid H_1,A\right)$, then we compare this maximum value against $P\left(\left\{|r(k)|\right\}_{k=0}^{N-1} \mid H_0\right)$.
By taking the logarithm, the GLRT simplifies to comparing the log-likelihood ratio to the threshold
\begin{align}
	\ln \Lambda = - \frac{N \hat{A}^2}{2\sigma^2} + \sum_{k=0}^{N-1} \ln I_0 \left( \frac{|r(k)| \hat{A}}{\sigma^2} \right) \underset{H_0}{\overset{H_1}{\gtrless}} 0
	\label{GLRT}.
\end{align}
To find $\hat{A}$, we take the derivative of $\ln \Lambda$ with respect to $A$ and set it to zero
\begin{align}
	\frac{d}{d\hat{A}} \ln \Lambda = - \frac{N \hat{A}}{\sigma^2} + \sum_{k=0}^{N-1} \frac{I_1 \left( \frac{|r(k)| \hat{A}}{\sigma^2} \right)}{I_0 \left( \frac{|r(k)| \hat{A}}{\sigma^2} \right)} \cdot \frac{|r(k)|}{\sigma^2} = 0
\end{align}
which yields
\begin{align}
	\hat{A}= \frac{1}{N} \sum_{k=0}^{N-1} \frac{|r(k)| I_1 \left( \frac{|r(k)| \hat{A}}{\sigma^2} \right)}{I_0 \left( \frac{|r(k)| \hat{A}}{\sigma^2} \right)}.
	\label{MLE}
\end{align}
Due to the nonlinear property of modified Bessel function of the first kind, we cannot directly derive a closed-form of $\hat{A}$. We will discuss the approximation of the $I_v(.)$ in the low SNR case and high SNR case separately to derive an analytic expression in the following parts.

\subsection{Low SNR Case} \label{subsec:low_snr}

For low SNR case where the argument $x$ of the modified Bessel functions of the first kind $I_v(x)$ is close to $0$, its approximation can be expressed as
\begin{align}
	I_0(x) &\approx 1 + \frac{x^2}{4} + \frac{x^4}{64}\\
	I_1(x) &\approx \frac{x}{2} + \frac{x^3}{16} + \frac{x^5}{384} .
\end{align}
The ratio $\frac{I_1(x)}{I_0(x)} $ can be reduced to
\begin{align}
	\frac{I_1(x)}{I_0(x)} \approx \frac{x}{2} - \frac{x^3}{16}.
\end{align}
Substituting this low SNR approximation into the fixed-point equation (\ref{MLE}), the MLE reduces to
\begin{align}
	\hat{A}_{\text{L}} &\approx \frac{1}{N} \sum_{k=0}^{N-1} |r(k)| \left[ \frac{\frac{\hat{A}_{\text{L}} |r(k)|}{2\sigma^2} - \frac{\left(\frac{\hat{A}_{\text{L}} |r(k)|}{\sigma^2}\right)^3}{16}}{1 + \frac{\left(\frac{\hat{A}_{\text{L}} |r(k)|}{\sigma^2}\right)^2}{4} + \frac{\left(\frac{\hat{A}_{\text{L}} |r(k)|}{\sigma^2}\right)^4}{64}} \right] \\
	&\approx \frac{1}{N} \sum_{k=0}^{N-1} |r(k)| \left[ \frac{\frac{\hat{A}_{\text{L}} |r(k)|}{2\sigma^2} - \frac{\hat{A}_{\text{L}}^3 |r(k)|^3}{16\sigma^6}}{1} \right]  \\
	&= \frac{\hat{A}_{\text{L}}}{2N\sigma^2} \sum_{k=0}^{N-1} |r(k)|^2 - \frac{\hat{A}_{\text{L}}^3}{16N\sigma^6} \sum_{k=0}^{N-1} |r(k)|^4.
\end{align}
Rearranging terms yields
\begin{align}
	\hat{A}_{\text{L}} \approx \frac{\hat{A}_{\text{L}} }{2\sigma^2}M_2 - \frac{\hat{A}_{\text{L}}^3 }{16\sigma^6}M_4
\end{align}
where the sample moments of $p$-th order are defined as
\begin{align}
	M_p \triangleq \frac{1}{N} \sum_{k=0}^{N-1} |r(k)|^p.
\end{align}
This leads to the cubic score equation in \( \hat{A}_{\text{L}} \):
\begin{align}
	\hat{A}_{\text{L}} \left( -1 + \frac{1}{2\sigma^2}M_2 - \frac{\hat{A}_{\text{L}}^2 }{16\sigma^6}M_4 \right) \approx 0.
\end{align}
Besides the boundary root \( \hat{A}_{\text{L}} = 0 \), the non-trivial low SNR MLE solves the quadratic in \( \hat{A}_{\text{L}}^2 \):
\begin{align}
	\hat{A}_{\text{L}}^2 = \frac{8 \sigma^4}{M_4} \left( M_2 - 2 \sigma^2 \right)_+
\end{align}
where $(x)_+ \triangleq \max\{x,0\}$ denotes the nonnegative part of $x$. Due to the natural boundary $ \hat{A}_{\text{L}} \geq 0 $, a non-negativity result is given by
\begin{align}
	\hat{A}_{\text{L}} = \sqrt{ \frac{8 \sigma^4}{M_4} \left(M_2 - 2 \sigma^2 \right)_+ }.
	\label{Alow}
\end{align}

Then we plug this $\hat{A}_{\text{L}}$ back into the GLRT in (\ref{GLRT}). Using the small-argument series of $I_0(x)$, the test statistic can be simplified to

\begin{align}
	\ln \Lambda 
	&\approx -\frac{N \hat{A}_{\text{L}}^2}{2\sigma^2} 
	+ \frac{N \hat{A}_{\text{L}}^2 M_2}{4\sigma^4} 
	- \frac{N \hat{A}_{\text{L}}^4 M_4}{64\sigma^8} \\[2mm]
	&= \frac{N\hat{A}_{\text{L}}^{2}}{4\sigma^{4}}\,(M_2-2\sigma^{2})
	\;-\; \frac{N\hat{A}_{\text{L}}^{4}M_4}{64\sigma^{8}}\\
	&\underset{H_0}{\overset{H_1}{\gtrless}}0.
	\label{eq:llr_group}
\end{align}
Using the result in (\ref{Alow}), we have
\begin{align}
	\ln \Lambda 
	&\approx \frac{N}{M_{4}}\,(M_{2}-2\sigma^{2})^{2}\\
	&\underset{H_0}{\overset{H_1}{\gtrless}}0.
	\label{eq:llr_final_form}
\end{align}
Due to the fact that $M_4 \geq 0$, the low SNR detector can be finally formed as
\begin{align}
T_{\text{E}}= M_2
\underset{H_0}{\overset{H_1}{\gtrless}} 2\sigma^2.
\label{Ed}
\end{align}

In the low SNR regime, the above analysis reveals that the optimal detection strategy simplifies into an energy detection scheme. Specifically, by approximating the modified Bessel function, We estimate the low SNR amplitude, and then the complex likelihood ratio test reduces to comparing the average received energy $T_\text{E}$ against the noise energy level \(2\sigma^2\). This outcome is significant: while energy detectors are often employed as the test statistic, here we first derive it rigorously as the optimal solution for minimizing the overall error probability at low SNR. Thus, rather than being a mere heuristic, the energy detector emerges naturally from the fundamental statistical properties of the problem, confirming its optimality in scenarios where the signal is weak relative to the noise.

\subsection{High SNR Case} \label{subsec:high_snr}
For $I_v(x)$, when the variable $x$ is very large, it can be expressed as 
\begin{align}
&	I_0(x) \approx \frac{e^x}{\sqrt{2\pi x}} \left[ 1 + \frac{1}{8x} \left( 1 + \frac{9}{2(8x)} \left( 1 +  \ldots \right) \right) \right] \\
&	I_1(x) \approx \frac{e^x}{\sqrt{2\pi x}} \left[ 1 - \frac{3}{8x} \left( 1 + \frac{5}{2(8x)} \left( 1 +  \ldots \right) \right) \right]
\end{align}
and thus the Bessel function ratio in (\ref{MLE}) can be approximated
\begin{equation}
	\frac{I_1(x)}{I_0(x)} \approx 1-\frac{4}{8x+1} \approx 1-\frac{1}{2x}
\end{equation}
With those approximations, the solution of the MLE for (\ref{MLE}) can be simplified as the solution of the following equation \cite{wu2021maximum}:
\begin{align}
	\hat{A}^2 - \frac{1}{N} \sum_{k=0}^{N-1} |r(k)|\hat{A} + \frac{\sigma^2}{2} = 0
	\label{squareAsolution} . 
\end{align}
Define the first order sample moment as
\begin{align}
	M = \frac{1}{N} \sum_{k=1}^{N} |r(k)|
\end{align}
and the MLE of amplitude is
\begin{align}
	\hat{A}=\frac{1}{2}M  \pm \frac{1}{2} \sqrt{M ^2-2\sigma^2} . 
	\label{ALA}
\end{align}

For large argument approximation where $M ^2 \gg 2\sigma^2$, the solution with the $-$ sign results in an estimate approaching zero. The solution with the $+$ sign is close to $M$ and is more realistic. Simulations show that, in the case of large noise, rare occasions with $M ^2-2\sigma^2 < 0$ occur, resulting in complex solutions to the quadratic equation in (\ref{ALA}). Hence, we propose to take the real part of (\ref{ALA}) as the amplitude estimator
\begin{equation}
	\hat{A}_{\text{H}} = \frac{1}{2} \left( M +  \left( \sqrt{ M^2 - 2\sigma^2 } \right)_+ \right).
	\label{ALAR}
\end{equation}
Now we can use the MLE result $\hat{A}_{\text{H}}$ to simplify the GLRT from (\ref{GLRT}). When the parameter $x$ is very large, the logarithm of $I_0(x)$ has an approximation
\begin{align}
	\nonumber
	\ln I_0 (x) &\approx \ln \left\{\frac{e^x}{\sqrt{2\pi x}} \left[ 1 + \frac{1}{8x} \left( 1 + \frac{9}{2(8x)} \left( 1 + \ldots \right) \right) \right] \right\} \\
	\nonumber
	&\approx x - \frac{1}{2} \ln(2\pi x) \\
	&\approx x
\end{align}
and thus the test statistic (\ref{GLRT}) can be simplified as 
\begin{align}
	\nonumber
	\ln \Lambda &= - \frac{N \hat{A}_{\text{H}}^2}{2\sigma^2} + \sum_{k=0}^{N-1} \ln I_0 \left( \frac{|r(k)| \hat{A}_{\text{H}}}{\sigma^2} \right) \\
	\nonumber
	&\approx - \frac{N \hat{A}_{\text{H}}^2}{2\sigma^2} + \sum_{k=0}^{N-1}\frac{|r(k)| \hat{A}_{\text{H}}}{\sigma^2}\\
	\nonumber
	&= - \frac{N \hat{A}_{\text{H}}^2}{2\sigma^2} + \frac{\hat{A}_{\text{H}}}{\sigma^2}\sum_{k=0}^{N-1}|r(k)| \\ &\underset{H_0}{\overset{H_1}{\gtrless}} 0 .
\end{align}
Using the result in (\ref{squareAsolution}), it can be further simplified as  
\begin{align}
		\nonumber
\ln \Lambda &=\frac{N }{2\sigma^2} \left(2\hat{A}_{\text{H}}\sum_{k=0}^{N-1}|r(k)| - \hat{A}_{H}^2\right)\\
\nonumber
&=\frac{N }{2\sigma^2} \left( \hat{A}_{\text{H}}^2 - \sigma^2 \right)\\
 &\underset{H_0}{\overset{H_1}{\gtrless}} 0 
\end{align}
which reduces to
\begin{align}
	 \hat{A}_{\text{H}} \underset{H_0}{\overset{H_1}{\gtrless}} \sigma
	 \label{AD} .
\end{align}

Here, we can further simplify (\ref{AD}) by analyzing the square root item of $\hat{A}_{\text{H}}$. Recall $\hat{A}_{\text{H}}$ in (\ref{ALAR}), when the inner square root part of $\hat{A}_{\text{H}}$ is real ($M^2 \geq 2\sigma^2$), we can show that the amplitude detector can be simplified as 
\begin{align}
	M \underset{H_0}{\overset{H_1}{\gtrless}} \frac{3}{2} \sigma
\end{align}
When $M^2 < 2\sigma^2$, the square root is imaginary, and we take the real part:
\begin{equation}
	\hat{A}_{\text{H}} = \frac{1}{2} M
\end{equation}
Since $M < \sqrt{2} \sigma$, $\hat{A}_{\text{H}} < \sigma$, so the decision is $H_0$.
Thus, the amplitude detector can be transformed into a threshold on $M$:
\begin{align}
	T_\text{A} = M \underset{H_0}{\overset{H_1}{\gtrless}} \frac{3}{2} \sigma .
	\label{Ad}
\end{align}

As shown in (\ref{Ad}), in the high SNR scenario, the derived optimal detector takes the form of an amplitude detector, which compares the sample mean of the signal amplitude directly against the noise standard deviation $\frac{3}{2}\sigma$. 

In summary, we first derive the optimum detection framework by working exclusively with the noisy received signal magnitudes $\left\{|r(k)|\right\}_{k=0}^{N-1}$ and employing a magnitude estimation. Our new detector circumvents the need for precise frequency alignment and complex baseband processing. Both the optimum detectors derived in low and high SNR stand in stark contrast to traditional IQ-based detection methods, which require phase synchronization and detailed frequency estimates. Our derivation proves that at low SNR, the energy detector provides the optimal solution, while at high SNR, the theoretically optimum detector turns out to be an amplitude detector.

\section{Performance Analysis}
\subsection{Performance Analysis of the Energy Detector}
In this section, we present the derivation of the false alarm rate $P_{\text{FA}}^{(E)} $ and misdetection rate $P_{\text{MD}}^{(E)} $ for the proposed energy detector to further calculate the probability of error $P_{\text{e}}^{(E)} $. The energy detector decides between the presence or absence of a signal based on the energy $T_\text{E}$. The detailed derivation includes distribution analysis under both hypotheses $H_0$ and $H_1$, and expresses $P_{\text{FA}}^{(E)} $, $P_{\text{MD}}^{(E)} $ and $P_{\text{e}}^{(E)} $.
\subsubsection{Probability of False Alarm}
For the energy detector in low SNR case, the null hypothesis \( H_0 \) is
\[
H_0: \quad r(k) = n(k), \quad k = 0, 1, \ldots, N-1
\]
and the test statistic $T_\text{E}$ under $H_0$ can be represented as
\begin{align}
	\nonumber
	T_\text{E}^{(0)} &= \frac{1}{N} \sum_{k=0}^{N-1} |r(k)|^2 \\
	\nonumber
	&= \frac{1}{N} \sum_{k=0}^{N-1} |n(k)|^2 \\
	&= \frac{1}{N} \sum_{k=0}^{N-1} \left( n_I^2(k) + n_Q^2(k) \right).
\end{align}
Since \( n_{\text{I}}(k) \) and \( n_{\text{Q}}(k) \) are independent Gaussian random variables with zero mean and variance \( \sigma^2 \), the sum \( n_{\text{I}}^2(k) + n_{\text{Q}}^2(k) \) follows an exponential distribution with mean \( 2\sigma^2 \). Therefore, \( T_\text{E}^{(0)} \) is the average of \( N \) independent exponential random variables, which follows a gamma distribution
\begin{align}
	T_\text{E}^{(0)} &\sim \text{Gamma}\left( N, \frac{2\sigma^2}{N} \right)
\end{align}
with shape parameter \( \alpha_E = N \) and scale parameter \( \theta_E = \frac{2\sigma^2}{N} \).
And we can calculate the mean and variance of \( T_\text{E} \) under \( H_0 \) as
\begin{align}
	\mathbb{E}\left[T_\text{E}^{(0)} \right] &= N \theta = 2\sigma^2 \\
	\mathrm{Var}\left(T_\text{E}^{(0)} \right) &= N \theta^2 = \frac{4\sigma^4}{N}.
\end{align}

The probability of false alarm is the probability that \( E \) exceeds the detection threshold \( 2\sigma^2 \) under \( H_0 \), which is
\begin{align}
	P_{\text{FA}}^{(E)} = 1 - F_{T_\text{E}^{(0)}}\left( 2\sigma^2 \right)
\end{align}
and the cumulative distribution function (CDF) of \( T_\text{E}^ {(0)} \) is the CDF of the gamma distribution
\begin{align}
	F_{T_\text{E}^{(0)}}(2\sigma^2) &= \frac{\gamma\left( N, N \right)}{\Gamma(N)}
\end{align}
where $\Gamma(N)$ is the Gamma function and \( \gamma(N, z) \) is the lower incomplete gamma function
\begin{align}
	\Gamma(N) &=  (N-1)! \\
	\gamma(N, N) &= \int_{0}^{N} t^{N-1} e^{-t} \, dt .
\end{align}
According to this, we can derive that the false alarm rate for ED is 
\begin{align}
	P_{\text{FA}}^{(E)} &= 1 - \frac{\gamma\left( N, N \right)}{\Gamma(N)}
	\label{func_pfa_ed} .
\end{align}

\subsubsection{Probability of Misdetection }

The signal under \( H_1 \) is
\[
H_1: \quad r(k) = A e^{j(\omega k + \theta)} + n(k), \quad k = 0, 1, \ldots, N-1
\]

We first analyse the distribution of $|r(k)|^2$.
The complex signal \(  A e^{j(\omega k + \theta)}  \) and noise \( n(k) \) can be decomposed into their real and imaginary components
\begin{align}
	s(k) &= A \cos(\omega k + \theta) + j A \sin(\omega k + \theta) \\
	n(k) &= n_{\text{I}}(k) + j n_{\text{Q}}(k)
\end{align}
where \( n_{\text{I}}(k) \) and \( n_{\text{Q}}(k) \) are independent zero-mean Gaussian random variables with variance \( \sigma^2 \).

Thus, the received signal can be written as
\begin{align}
	r(k) &= \left[ A \cos(\omega k + \theta) + n_{\text{I}}(k) \right] + j \left[ A \sin(\omega k + \theta) + n_{\text{Q}}(k) \right]
\end{align}
with means
	\begin{align}
		\mu_{\text{I}} &= A \cos(\omega k + \theta) \\
		\mu_{\text{Q}} &= A \sin(\omega k + \theta)
	\end{align}
and variances $\sigma^2$. We can derive that
\begin{align}
	|r(k)|^2 &= [r_{\text{I}}(k)]^2 + [r_{\text{Q}}(k)]^2 .
\end{align}
This expression represents the sum of the squares of two independent Gaussian random variables $Z_{\text{I}}$ and ${Z_{\text{Q}}}$ with non-zero means and equal variances:
\begin{align}
	\mu_{Z_{\text{I}}} &= \frac{\mu_I}{\sigma} = \frac{A \cos(\omega k + \theta)}{\sigma} \\
	\mu_{Z_{\text{Q}}} &= \frac{\mu_Q}{\sigma} = \frac{A \sin(\omega k + \theta)}{\sigma} .
\end{align}
Thus, the $|r(k)|^2$ follows a non-central chi-squared distribution with 2 degrees of freedom and the non-centrality parameter: 
\begin{align}
	\lambda_\text{E} &= \mu_{Z_{\text{I}}}^2 + \mu_{Z_{\text{Q}}}^2 = \frac{A^2}{\sigma^2}
\end{align}
and we can conclude that
\begin{align}
	\left( \frac{r_{\text{I}}(k)}{\sigma} \right)^2 + \left( \frac{r_{\text{Q}}(k)}{\sigma} \right)^2 &\sim \chi'^2_2\left( \lambda_\text{E} \right) 
\end{align}
and thus
\begin{align}
	[r_{\text{I}}(k)]^2 + [r_{\text{Q}}(k)]^2 &= |r(k)|^2 \sim \sigma^2 \chi'^2_2\left( \lambda_\text{E} \right)
	\label{chisquaredis} .
\end{align}

Then, we can derive the distribution of test statistic \( T_\text{E} \) under \( H_1 \) as
\begin{align}
	\nonumber
	T_\text{E}^{(1)} &= \frac{1}{N} \sum_{k=0}^{N-1} \left| A e^{j(\omega k + \theta)} + n(k) \right|^2 \\
	&= \frac{1}{N} \sum_{k=0}^{N-1} \left( A^2 + 2 A \operatorname{Re} \left\{ e^{-j(\omega k + \theta)} n(k) \right\} + |n(k)|^2 \right) .
\end{align}

Since \( n(k) \) is complex Gaussian noise with zero mean, the cross-term $ \operatorname{Re} \left\{ e^{-j(\omega k + \theta)} n(k) \right\}$ has zero mean and variance \( \sigma^2 \).
Therefore, the expected value of $T_\text{E}^{(1)} $ can be derived as
\begin{align}
	\mathbb{E}\left[T_\text{E}^{(1)} \right] &= A^2 + 2\sigma^2 .
\end{align}

To find the distribution of \( T_\text{E} \) under \( H_1 \), we have $|r(k)|^2 \sim \sigma^2 \chi'^2_2\left( \lambda_\text{E} \right)$ from (\ref{chisquaredis}). Therefore, the sum $\sum_{k=0}^{N-1} |r(k)|^2 $
follows a non-central chi-squared distribution with \( 2N \) degrees of freedom and non-centrality parameter \( \Lambda_E^{(1)} = N \lambda_\text{E} = \dfrac{NA^2}{\sigma^2} \): 
\begin{align}
	\sum_{k=0}^{N-1} |r(k)|^2 \sim \sigma^2 \cdot \chi'^2_{2N}\left( \dfrac{NA^2}{\sigma^2} \right). 
\end{align}
The test statistic $T_\text{E}^{(1)} = \frac{1}{N}\sum_{k=0}^{N-1} |r(k)|^2 $, so we have
\begin{align}
	T_\text{E}^{(1)} &\sim \frac{\sigma^2}{N} \cdot \chi'^2_{2N}\left( \dfrac{NA^2}{\sigma^2} \right). 
\end{align}
The mean and variance of $T_\text{E}^{(1)}$ are:
\begin{align}
	\mathbb{E}\left[T_\text{E}^{(1)} \right] &= \frac{\sigma^2}{N} \left(2N+\dfrac{NA^2}{\sigma^2} \right)=\sigma^2 + A^2 \\
	\mathrm{Var}\left(T_\text{E}^{(1)} \right) &= 2\frac{\sigma^4}{N^2} \left(2N+2\dfrac{NA^2}{\sigma^2} \right)=\frac{4\sigma^4 + 4\sigma^2 A^2}{N}. 
\end{align}

The probability of misdetection is the cumulative distribution function of the non-central chi-squared distribution that \( T_\text{E}^{(1)} \) falls below the threshold \( 2\sigma^2 \)
\begin{align}
	P_{\text{MD}}^{(E)} &= F_{T_\text{E}^{(1)}}\left( 2\sigma^2 \mid H_1 \right)\\
	&= 1 - Q_{N}\left( \frac{A}{\sigma}\sqrt{N}, \sqrt{2N} \right)
	\label{PMDED}
\end{align}
where \( Q_N(a, b) \) is the generalized Marcum Q-function
\begin{align}
	Q_N(a, b) &= \int_b^\infty x \left( \frac{x}{a} \right)^{N-1} \exp\left( -\frac{x^2 + a^2}{2} \right) I_{N-1}(a x) \, dx.
	\label{func_pmd_ed}
\end{align}

Thus, according to (\ref{func_pfa_ed}) and (\ref{PMDED}), the probability of error for energy detector is 
\begin{align}
	\nonumber
	P_{\text{e}}^{(E)} &= P_{\text{MD}}^{(E)} P(H_1) + P_{\text{FA}}^{(E)} P(H_0)\\
	&= \frac{1}{2} \left( 2 - \frac{\gamma\left( N, N \right)}{\Gamma(N)} - Q_{N}\left( \frac{A}{\sigma}\sqrt{N}, \sqrt{2N} \right)\right)
\end{align}

\subsection{Performance Analysis of the Amplitude Detector}
In this section, we presents the derivation of the false alarm rate $P_{\text{FA}}^{(A)}$ and misdetection rate $P_{\text{MD}}^{(A)}$ for the proposed amplitude detector to further calculate the probability of error $P_{\text{e}}^{(A)}$. The amplitude detector decides between the presence or absence of a signal based on the estimated amplitude $T_\text{A}$ calculated from received signal magnitudes. The detailed derivation includes approximations under both hypotheses $H_0$ and $H_1$, and expresses $P_{\text{FA}}^{(A)}$, $P_{\text{MD}}^{(A)}$ and $P_{\text{e}}^{(A)}$ using appropriate statistical distributions.

\subsubsection{Probability of False Alarm}
Under hypothesis $H_0$, $r(k) = n(k)$, the magnitude $|r(k)|$ follows a Rayleigh distribution with the probability density function (PDF):
	\begin{equation}
		f_{|r|}(x) = \frac{x}{\sigma^2} e^{- \frac{x^2}{2\sigma^2}}, \quad x \geq 0.
	\end{equation}
We consider the sum of the magnitudes over \(N\) samples under $H_0$:
	\begin{equation}
		S_0 = NT_\text{A}^{(0)} = \sum_{k=1}^{N} |n(k)|.
	\end{equation}
Our goal is to find the CDF of the test statistic $S_0$ or $M$ and thus calculate the outage probability. The sum \(S_0\) of Rayleigh-distributed random variables does not have a simple closed-form PDF. However, in \cite{hu2005accurate}, a modified CDF approximation for the sum of Rayleigh random variables is given.
\begin{align}
	F_{S_0}(x) = F_{\text{SAA}}(x) - F_{\text{error}}(x)
\end{align}
where $F_{\text{SAA}}(t)$ is the small argument approximation (SAA) for the sum CDF
\begin{align}
	F_{\text{SAA}}(x) &= 1 - e^{-\frac{x^2}{2bN\sigma^2}} \sum_{k=0}^{N-1} \frac{\left( \frac{x^2}{2bN\sigma^2} \right)^k}{k!}\\
	b &= \frac{\sigma^2}{N} \left( (2N - 1)!! \right)^{1/N}
\end{align}
where $(2N - 1)!!=(2N-1)\times(2N-3)\times\ldots\times3\times1$.

The difference item $F_{\text{error}}(x)$ is an empirically difference from the SAA approximation to the real value, which is
\begin{small}
\begin{align}
	F_{\text{error}}(x) \approx \frac{a_0 \cdot x \left( x - a_2 \sigma \sqrt{N} \right)^{2N-1} \exp\left({-\frac{a_1 \left( x - a_2 \sigma \sqrt{N} \right)^2}{2 \sigma^2 N b}}\right)} {\left( \sigma \sqrt{N} \right)^{2N} \cdot 2^{N-1} \left( \frac{b}{a_1} \right)^N N!}	
\end{align}
\end{small}where the coefficients $a_0, a_1, a_2$ are calculated by fitting the error function to the difference between the Rayleigh sum variables' CDF and the $F_\text{SAA}$.

Thus, the $P_{{\text{FA}}}$ is defined as the probability that the sum $S_0$ exceeds the threshold under $H_0$ is given by
\begin{align}
	\nonumber
	P_{\text{{FA}}}^{(A)} &= 1 - F_{S_0}\left( \frac{3}{2} N \sigma \right)\\
&= 1 - \left[ F_{\text{SAA}}\left( \frac{3}{2} N \sigma \right) - F_{\text{error}}\left( \frac{3}{2} N \sigma \right) \right] .
\label{PFAAD}
\end{align}

	\subsubsection{Probability of Misdetection}
Under hypothesis $H_1$, $r(k) = A e^{j(\omega k + \theta)} + n(k)$. The magnitude $|r(k)|$ follows a Rician distribution with PDF
	\begin{equation}
		f_{|r|}(x) = \frac{x}{\sigma^2} e^{- \frac{x^2 + A^2}{2\sigma^2}} I_0\left( \frac{x A}{\sigma^2} \right), \quad x \geq 0,
	\end{equation}
The misdetection rate is denoted by
\begin{align}
	P_{\text{{MD}}}^{(A)} &= P\left( S_1 < \frac{3}{2} N \sigma \mid H_1 \right) .
\end{align}
where 
\begin{align}
	S_1 = NT_\text{A}^{(1)} = \sum_{k=1}^{N} |A e^{j(\omega k + \theta)} + n(k)|.
\end{align} 
However, the Rician sum random variables also don't have a closed-form representation. From \cite{garcia2021novel}, the CDF for the sum of $N$ i.i.d. Rician random variables $S_1$ is given by
\begin{align}
	F_{S_1}(r) &= \frac{1}{\sigma^{2N}} \exp\left( -\frac{\nu^2 N}{2 \sigma^2} \right) \sum_{i=0}^{\infty} \frac{\delta_i r^{2(i+N)}}{\Gamma(2(i + N) + 1)}
\end{align}
where $\nu = A/\sigma$ is the Rice factor and $\delta_i$ are recursively calculated coefficients computed recursively by
\begin{small}
\begin{align}
	\delta_0 &= 1 \\
	\delta_i &= \frac{1}{i} \sum_{k=1}^i (N k + k - i) \delta_{i - k} \sum_{l=0}^k \frac{\Gamma(2(k + 1))}{(2 \sigma^2)^l \left( \frac{\nu}{2\sigma^2} \right)^{2(k - l)} (k - l)!^2 \, l!}
\end{align}
\end{small}
To find $P_{\text{{MD}}}^{(A)}$, we substitute the AD threshold $r = \frac{3}{2} N \sigma$ in $F_{S_1}(r)$:
\begin{align}
	\nonumber
	P_{\text{MD}}^{(A)} &= F_{S_1}\left( \frac{3}{2} N \sigma \right) \\
	&= \frac{1}{\sigma^{2N}} \exp\left( -\frac{\nu^2 N}{2 \sigma^2} \right) \sum_{i=0}^{\infty} \frac{\delta_i \left( \frac{3}{2} N \sigma \right)^{2(i+N)}}{\Gamma(2(i + N) + 1)}
	\label{P_md_ad_approx} .
\end{align}

Thus, based on the false alarm rate and the misdetection rate given in  (\ref{PFAAD}) and (\ref{P_md_ad_approx}), the probability of error for amplitude detector is 
\begin{align}
	\nonumber
	P_{\text{e}}^{(A)} = & \, P_{\text{MD}}^{(A)} P(H_1) + P_{\text{FA}}^{(A)} P(H_0) \\
		\nonumber
	= & \, \frac{1}{2} \left( 1 - \left[ F_{\text{SAA}}\left( \frac{3}{2} N \sigma \right) - F_{\text{error}}\left( \frac{3}{2} N \sigma \right) \right] \right. \\
	&+ \left. \frac{1}{\sigma^{2N}} \exp\left( -\frac{\nu^2 N}{2 \sigma^2} \right) \sum_{i=0}^{\infty} \frac{\delta_i \left( \frac{3}{2} N \sigma \right)^{2(i+N)}}{\Gamma(2(i + N) + 1)} \right) .
\end{align}

\section{Reliability-based intelligent decision rule}
\label{rid}

In this section, we propose a reliability-based intelligent decision rule for our optimal detectors. 
In the previous sections, we have shown that, based on the received magnitudes, the energy detector and amplitude detector naturally arise as the optimal detectors in different SNR regimes.
Each detector is well justified in its own regime, but neither alone is
uniformly reliable across all SNRs. In particular, around the
transition region where the low and high SNR approximations intersect,
ED and AD can exhibit noticeably different error behaviors and may even
produce conflicting decisions for the same observation.

A naive solution is to \emph{hard switch} between ED and AD at a fixed
design SNR. However, such a global switch point:
(i) depends on unknown channel conditions,
(ii) ignores the actual observed statistics in each realization, and
(iii) may cause performance loss when the true SNR deviates from the
design value.

Our goal is therefore to design a \emph{reliability-based intelligent
	decision} rule that:
\begin{itemize}
	\item operates on the same low-complexity statistics already computed
	for ED and AD;
	\item retains the good behavior of ED at low SNR and AD at high SNR;
	\item adaptively resolves disagreements between ED and AD on a
	per-observation basis, favoring the locally more reliable decision
	quantified via their error probabilities.
\end{itemize}

The RID rule will use only $(\hat H_\text{E}, \hat H_\text{A})$
and simple functions of $(T_\text{E}, T_\text{A})$ to produce a final
decision, without introducing new front-end complexity.

\subsection{RID algorithm}
The RID algorithm begins by specifying the required inputs and proceeds to compute the necessary statistics from the received signal magnitudes. The inputs consist of the sequence of magnitudes $\{|r(k)|\}_{k=1}^{N}$ and the known noise variance $\sigma^2$.

Based on the proposed optimal ED and AD, we obtain the local decisions $(\hat H_\text{E}, \hat H_\text{A})$ respectively
\begin{align}
	\nonumber
	T_{\text{E}} \underset{H_0}{\overset{H_1}{\gtrless}} 2\sigma^2 &\;\; \Longrightarrow\;\;   \hat H_\text{E}\\
	\nonumber
	T_\text{A} \underset{H_0}{\overset{H_1}{\gtrless}} \frac{3}{2} \sigma  &\;\; \Longrightarrow\;\;   \hat H_\text{A}.
\end{align}

Our intelligent decision rule is proposed to fuse the local decisions $(\widehat{H}_\text{E},\widehat{H}_\text{A})$ into a final decision $\widehat{H}_\text{RID}$. The fusion logic is entirely
determined by whether the two local decisions agree or disagree, and it is
driven by the reliability of each detector as quantified by its false alarm
and misdetection probabilities derived in Section~III. The
RID keeps the consensus decision when ED and AD agree, and in the case of a disagreement it selects the hypothesis whose local decision is statistically more
reliable for the given operating point.

The different decision patterns handled by the RID can be summarized as:
\begin{itemize}
	\item \textbf{Agreement case:}
	\begin{itemize}
		\item \emph{Case (1): Both decide $H_1$:}
		\[
		\widehat{H}_\text{E} = \widehat{H}_\text{A} = H_1
		\;\Longrightarrow\;
		\widehat{H}_\text{RID} = H_1.
		\]
		This case occurs at sufficiently high SNR, where both detectors confidently detect the signal.
		\item \emph{Case (2): Both decide $H_0$:}
		\[
		\widehat{H}_\text{E} = \widehat{H}_\text{A} = H_0
		\;\Longrightarrow\;
		\widehat{H}_\text{RID} = H_0.
		\]
		This case occurs at very low SNR, where both detectors favor the noise only hypothesis.
	\end{itemize}
	
	\item \textbf{Disagreement case: reliability-based tie-breaking}
	\begin{itemize}
		\item \emph{Case (3): ED decides $H_1$, AD decides $H_0$:}
		
			According to the false alarm tail under $H_0$ at the ED threshold, the likelihood reliability of decision is given by
			\begin{align}
				P_E\left( T_{\text{E}}>2\sigma^2 \mid H_0\right)
				&= P_{\text{FA}}^{(E)}
				= 1 - \frac{\gamma(N,N)}{\Gamma(N)} .
			\end{align}
			Similarly, we use the misdetection tail under $H_1$ at the AD threshold, the likelihood reliability of decision is given by
			\begin{align}
				P_A\left(T_{\text{A}} \le \frac{3}{2}\sigma \mid H_1\right)
				&= P_{\text{MD}}^{(A)}
				= F_{S_0}\left(\frac{3}{2}N\sigma\right).
			\end{align}
			By comparing the decision reliability, the decision rule for this case is given by
			\begin{align}
				P_{\text{FA}}^{(E)} \underset{H_1}{\overset{H_0}{\gtrless}} P_{\text{MD}}^{(A)} .
			\end{align}
		\item \emph{Case (4): ED decides $H_0$, AD decides $H_1$:}
		
				We use the misdetection tail under $H_1$ at the ED threshold, the likelihood reliability of decision is given by
				\begin{align}
					\nonumber
					P_E\left(T_{\text{E}}\le 2\sigma^2\mid  H_1\right)
					&= P_{\text{MD}}^{(E)}\\
					&= 1 - Q_{N}\left(\frac{A}{\sigma}\sqrt{N}, \sqrt{2N}\right)
				\end{align}
				and we can use the false alarm tail under $H_0$ at the AD threshold to represent the likelihood reliability of decision
				\begin{align}
					P_A\left( T_{\text{A}} > \frac{3}{2}\sigma\mid H_0\right)
					&= P_{\text{FA}}^{(A)}
					= 1 - F_{S_1}\left(\frac{3}{2}N\sigma\right)
				\end{align}
				and the decision rule for case (4) is 
				\begin{align}
					P_{\text{MD}}^{(E)} \underset{H_0}{\overset{H_1}{\gtrless}} P_{\text{FA}}^{(A)}  .
				\end{align}
	\end{itemize}
\end{itemize}

However, from an implementation perspective, the exact value of $A$ is unknown, and the corresponding reliability likelihoods is hard to be evaluated explicitly at the receiver. Therefore, a simplified decision rule is required for the disagreement cases of the RID detector. In the following, we develop such a rule, which enables the RID decision to be realized directly from the received amplitudes.

\subsection{Simplified Decision Rule for Case (3): }
\label{subsec:RID_case3}
In the disagreement case where the ED decides $H_1$ while the AD decides $H_0$, we have

\begin{align}
	\nonumber
	T_{\text{E}} > 2\sigma^2 &\;\; \Longrightarrow\;\;   \hat H_\text{E} = H_1  \\
	\nonumber
	T_\text{A} < \frac{3}{2} \sigma  &\;\; \Longrightarrow\;\;   \hat H_\text{A} = H_0 .
\end{align}

Following the performance analysis in Section~III, the two reliability to be compared in this case are the ED false alarm tail under $H_0$ and the AD misdetection tail under $H_1$. The decision rule selects the hypothesis whose local decision is \emph{more reliable}, i.e., the one with the smaller error probability
\begin{align}
	P_{\text{FA}}^{(E)}(N) \underset{H_1}{\overset{H_0}{\gtrless}} P_{\text{MD}}^{(A)}(N,A).
	\label{eq:case3_basic_rule}
\end{align}

From the performance analysis, $P_{\text{MD}}^{(A)}(N,A)$ is strictly decreasing in $A$ for fixed $N$, whereas $P_{\text{FA}}^{(E)}(N)$ is a constant for a given $N$. Hence, there exists a unique switch point $A_\star(N)>0$ such that
\begin{align}
	P_{\text{MD}}^{(A)}\left(N,A_\star(N)\right) 
	= 
	P_{\text{FA}}^{(E)}(N).
	\label{eq:Astar_def}
\end{align}
Therefore, the comparison in \eqref{eq:case3_basic_rule} reduces to a one-side test on amplitude:
\begin{align}
	\widehat{H}_\text{RID} =
	\begin{cases}
		H_1, & \text{if } \hat{A} \ge A_\star^{(3)}(N),\\
		H_0, & \text{if } \hat{A} < A_\star^{(3)}(N),
	\end{cases}
	\label{eq:case3_Ahat_rule}
\end{align}

In this case, $H_1$ is favored by the ED, so we adopt the corresponding low SNR amplitude estimator $\hat{A}_{\text{L}}$ as the estimate of $A$. For each $N$, the threshold $A_\star^{(3)}(N)$ can be computed numerically offline using the expressions derived in the performance analysis section. Representative values of $A_\star^{(3)}(N)$ are listed in Table~\ref{tab:case3_breakpoints}. At runtime, the case~(3) decision then requires only a comparison between the estimated amplitude and the precomputed threshold in \eqref{eq:case3_Ahat_rule}, without any probability evaluations.


\begin{table}[htbp]
	\caption{Threshold of Estimated Amplitudes for Case~(3)\label{tab:case3_breakpoints}}
	\centering
	\begin{tabular}{|c||c|c|c|c|c|c|}
		\hline
		$N$ & 1 & 2 & 4 & 8 & 12 & 16 \\
		\hline
		$A_\star^{(3)} / \sigma$ & 1.414 & 1.202 & 1.059 & 0.986 & 0.961 & 0.948 \\
		\hline
	\end{tabular}
\end{table}
\subsection{Simplified Decision Rule for Case (4): }
Similarly, in the disagreement case where the ED decides $H_0$ while the AD decides $H_1$, we have the local decision

\begin{align}
	\nonumber
	T_{\text{E}} < 2\sigma^2 &\;\; \Longrightarrow\;\;   \hat H_\text{E} = H_0  \\
	\nonumber
	T_\text{A} > \frac{3}{2} \sigma  &\;\; \Longrightarrow\;\;   \hat H_\text{A} = H_1 .
\end{align}

In this case, the two reliability measures to be compared are the ED misdetection tail under $H_1$ and the AD false alarm tail under $H_0$. The RID rule is therefore

\begin{align}
	P_{\text{MD}}^{(E)}(N)  \underset{H_0}{\overset{H_1}{\gtrless}} P_{\text{FA}}^{(A)}(N,A).
	\label{eq:case4_basic_rule}
\end{align}

Also, for fixed observations $N$, $P_{\text{MD}}^{(E)}(N,A)$ is strictly decreasing in $A$, whereas $P_{\text{FA}}^{(A)}(N)$ is a constant that does not depend on $A$. Hence, there exists a unique breakpoint $A_\star^{(4)}(N)>0$ such that the comparison in \eqref{eq:case4_basic_rule} reduces to a one-sided test on the amplitude estimate
\begin{align}
	\widehat{H}_\text{RID} =
	\begin{cases}
		H_1, & \text{if } \hat{A} \ge A_\star^{(4)}(N),\\
		H_0, & \text{if } \hat{A} < A_\star^{(4)}(N),
	\end{cases}
	\label{eq:case4_Ahat_rule}
\end{align}
where the $A_\star^{(4)}(N)$ can be precomputed offline. In this case, $H_1$ is favored by the AD, which corresponds to the high-SNR detector. Hence, we employ the high-SNR amplitude estimator $\hat{A}_\text{H}$ based on $T_\text{A}$.

Moreover, since there is a one-to-one mapping between $T_\text{A}$ and $\hat{A}_\text{H}$ (refer to  \eqref{squareAsolution}), which is strictly increasing for $\hat{A}_\text{H}>0$, we can write

\begin{align}
	T_\text{A}
	= \hat{A} + \frac{\sigma^2}{2\hat{A}}.
	\label{eq:R_from_Ahat_case4}
\end{align}

Thus, the RID rule for case~(4) can be expressed as an equivalent test in terms of $T_\text{A}$. Define the precomputable threshold
\begin{align}
	M_\star^{(4)}(N,\sigma) 
	\triangleq 
	A_\star^{(4)}(N) + \frac{\sigma^2}{2A_\star^{(4)}(N)}.
	\label{eq:rstar4_def}
\end{align}
and the amplitude test of case (4) can be simplified to
\begin{align}
	\widehat{H} =
	\begin{cases}
		H_1, & \text{if } T_\text{A} \ge M_\star^{(4)}(N,\sigma),\\
		H_0, & \text{if } T_\text{A} < M_\star^{(4)}(N,\sigma).
	\end{cases}
	\label{eq:case4_R_rule}
\end{align}

We also provide some representative values of $A_\star^{(3)}(N)$ and $M_\star^{(4)}(N,\sigma)$ in Table~\ref{tab:case4_breakpoints}.

\begin{table}[htbp]
	\caption{Threshold of Estimated Amplitudes for Case~(4)\label{tab:case4_breakpoints}}
	\centering
	\begin{tabular}{|c||c|c|c|c|c|c|}
		\hline
		$N$ & 1 & 2 & 4 & 8 & 12 & 16 \\
		\hline
		$A_\star^{(4)} / \sigma$ & 1.414 & 1.230 & 1.093 & 1.021 & 0.995 & 0.984 \\
		\hline
		$M_\star^{(4)} / \sigma$ & 1.768 & 1.637 & 1.550 & 1.511 & 1.498 & 1.492 \\
		\hline
	\end{tabular}
\end{table}

\subsection{Performance Properties of the RID}
In this subsection, we now discuss several structural and performance properties of the proposed
RID detector.

\subsubsection*{1) Impossibility of Case~(4)}
Recall that $T_\text{E} = \tfrac{1}{N}\sum_{k=1}^{N} |r_k|^2$ and
$T_\text{A} = \tfrac{1}{N}\sum_{k=1}^{N} |r_k|$. By Jensen’s inequality (or
Cauchy–Schwarz), we have
\begin{align}
T_\text{E} \;=\; \frac{1}{N}\sum_{k=1}^{N} |r_k|^2
\;\ge\;
\Big( \frac{1}{N}\sum_{k=1}^{N} |r_k| \Big)^2
\;=\; T_\text{A}^2.
\end{align}
If the AD decides $H_1$, i.e., $\widehat{H}_\text{A} = H_1$,
then $T_\text{A} > \tfrac{3}{2}\sigma$. Combining these relations yields
\begin{align}
T_\text{E} \;\ge\; T_\text{A}^2
\;>\; \Big(\tfrac{3}{2}\sigma\Big)^2
\;=\; 2.25\sigma^2
\;>\; 2\sigma^2
\;=\; T_\text{E,th}
\end{align}
which forces the ED to decide $H_1$ as well, i.e.,
$\widehat{H}_\text{E} = H_1$. Consequently, the disagreement configuration
\[
\{\widehat{H}_\text{E} = H_0,\ \widehat{H}_\text{A} = H_1\}
\]
cannot happen under the chosen thresholds, and therefore Case~(4) cannot be observed in practice. Nevertheless, for other operating points (e.g., when the Bayesian prior probabilities of
$H_0$ and $H_1$ are incorporated into the threshold design), Case~(4) is possible to occur. Hence, the analysis of Case~(4) remains meaningful and of independent interest.

\subsubsection*{2) Impact on misdetection probability}
\label{ridmd}
An immediate consequence of the above structural property is that, under
$H_1$, the only possible disagreement between ED and AD is Case~(3),
namely $\{\widehat{H}_\text{E} = H_1,\ \widehat{H}_\text{A} = H_0\}$.
On this disagreement set, the RID uses the reliability-based rule
developed in Section~\ref{subsec:RID_case3} and flips part of the AD
decisions from $H_0$ to $H_1$ whenever the estimated amplitude exceeds
the switch point $A_\star^{(3)}(N)$.

Every such flip removes a miss event of the AD, while outside the
disagreement region the RID decision coincides with that of the AD.
Therefore, for every $A>0$ and fixed $(N,\sigma)$, the miss probability
of RID is never larger than that of AD, i.e.,
\begin{equation}
	P^{{(R)}}_{\text{MD}}(N,A/\sigma)
	\;\le\;
	P^{{(A)}}_{\text{MD}}(N,A/\sigma).
	\label{eq:md_property_RID}
\end{equation}
This inequality formalizes the intuition that, under $H_1$, the RID
rule exploits the additional reliability information offered by ED
without degrading the miss performance relative to AD.

\subsubsection*{3) Impact on false alarm probability}
Under $H_0$, the same ordering argument $T_\text{E} \ge T_\text{A}^2$ implies that the event
$\{\widehat{H}_\text{A} = H_1\}$ is contained in
$\{\widehat{H}_\text{E} = H_1\}$. Hence, any false alarm of the AD must also be a false alarm of the ED.
Moreover, the RID does not simply adopt all ED false alarms: it
declares $H_1$ only on a subset of $\{\widehat{H}_\text{E} = H_1\}$ for
which the estimated amplitude exceeds the switch threshold
$A_\star^{(3)}(N)$.

As a result, the false alarm probability of RID is lower bounded by that
of AD and upper bounded by that of ED:
\begin{equation}
	P^{({A})}_{\text{FA}}(N)
	\;\le\;
	P^{({R})}_{\text{FA}}(N)
	\;\le\;
	P^{({E})}_{\text{FA}}(N).
	\label{eq:fa_property_RID}
\end{equation}
The lower bound reflects that RID cannot reduce false alarms below those of
AD without sacrificing the reliability-based flipping rule, whereas the
upper bound shows that RID never worsens the false alarm performance beyond
that of the ED.

\subsubsection*{4) Tradeoff and qualitative behavior}
Combining \eqref{eq:md_property_RID} and \eqref{eq:fa_property_RID}, we see
that the RID rule trades a controlled increase in false alarms (bounded
by ED) for a guaranteed decrease in misses relative to AD. This tradeoff is
particularly beneficial in the SNR regime where ED and AD exhibit comparable
performance and their operating points cross. In that regime, the reduction
in miss probability tends to dominate, leading to an overall lower total
error probability
\[
P_e^{({R})} \;=\; \tfrac{1}{2}
\big(P_{\text{FA}}^{({R})} + P_{\text{MD}}^{({R})}\big)
\]
for RID compared to AD. At sufficiently high SNR, where ED and AD decisions
almost always agree, the RID output coincides with that of AD, and the
additional combining step incurs essentially no performance penalty.
These qualitative behaviors will be corroborated by the numerical results
in Section~\ref{sec:numerical}.

\section{Numerical Results and Performance Discussion}
\label{sec:numerical}

In this section, we validate the analytical expressions derived for the
ED, AD, and the proposed RID detector and discuss their
performance trends across a broad SNR range and different observation
lengths.

\subsection{Simulation Setup}
We consider the signal model in Section~II and evaluate the false alarm
probability $P_{\text{FA}}$, the misdetection probability
$P_{\text{MD}}$, and the total error probability
$P_{e} = \frac{1}{2}\left(P_{\text{FA}} + P_{\text{MD}}\right)$ as
functions of the SNR and the number of samples $N$. To isolate the
effect of detector design, the noise variance is normalized to
$\sigma^2 = 1$, and all Monte Carlo curves are obtained from
$10^{8}$ independent trials per operating point.

For the ED, the theoretical $P_{\text{FA}}^{(E)}$ and
$P_{\text{MD}}^{(E)}$ are computed from the closed-form expressions in
(51) and (71), respectively.
For the AD, $P_{\text{FA}}^{(A)}$ is evaluated using the Rayleigh-sum
approximation developed in Section~III-B, while
$P_{\text{MD}}^{(A)}$ is obtained via a semi-analytical approach that
uses the empirical CDF of the Rician sum statistic to avoid numerical
overflow in the high-SNR regime.
For the RID detector, we implement the simplified case-based rules in
Section~IV, with the switching thresholds $A_\star^{(3)}(N)$ (and $A_\star^{(4)}(N)$/$M_\star^{(4)}(N)$) precomputed offline from the reliability equalities.
All curves are shown on a $\log_{10}$ scale to
highlight the low-probability region of interest.

\subsection{False-Alarm Probability}
\begin{figure}[htbp]
	\centering
	\subfloat[$P_{\text{FA}}$ vs SNR ($N=16$).]{
		\includegraphics[width=0.95\linewidth]{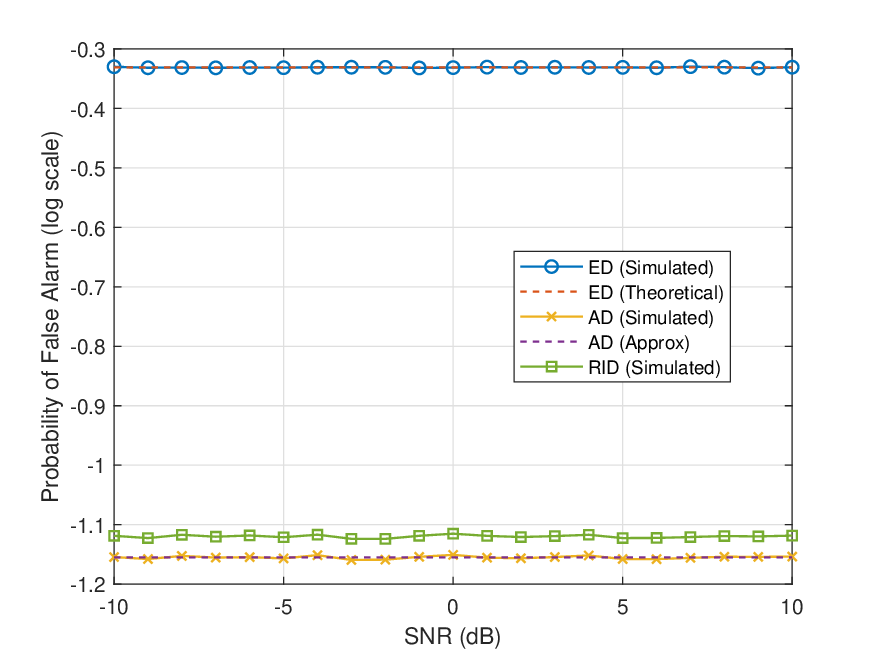}
		\label{fig:Pfa_snr}
	}
	
	\vspace{0.1cm} 
	\par
	
	\subfloat[$P_{\text{FA}}$ vs $N$.]{
		\includegraphics[width=0.95\linewidth]{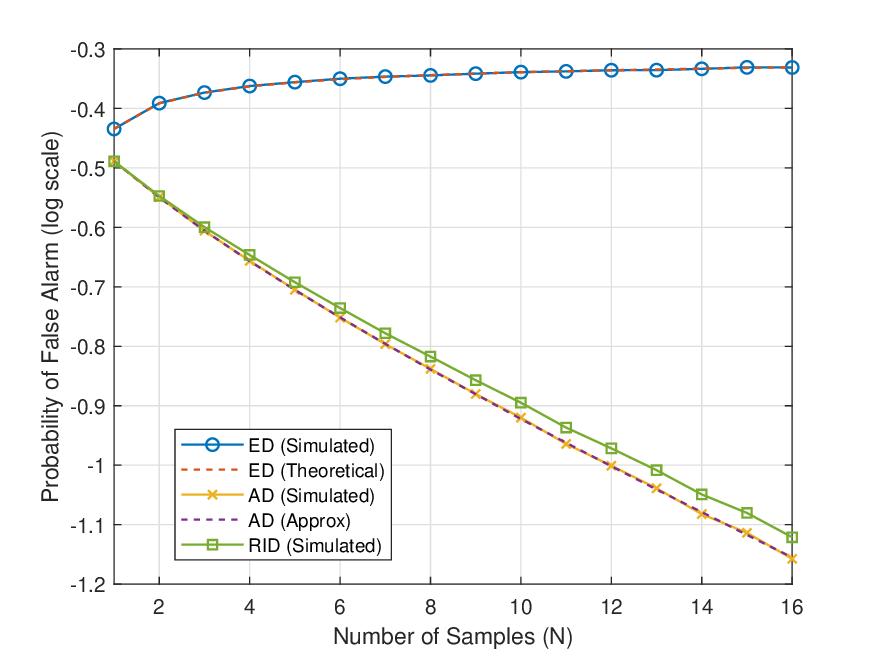}
		\label{fig:Pfa_N}
	}
	
	\caption{Simulated and analytical false alarm probability of ED, AD, and RID.}
	\label{fig:Pfa_all}
\end{figure}

Fig.~\ref{fig:Pfa_all} summarizes the false alarm performance for the detectors.
In Fig. \ref{fig:Pfa_all}-(a), $P_{\text{FA}}$ is plotted versus SNR
for a representative blocklength (e.g., $N=16$). The simulated curves
for both ED and AD are indistinguishable from their theoretical
predictions, confirming the accuracy of the analytical derivations.
As expected, the ED exhibits an almost SNR-independent false alarm
level, since its threshold is set solely from the noise statistics
under $H_0$ and does not depend on the signal amplitude. The AD also
shows negligible SNR dependence once the threshold is fixed, but its
false alarm probability is substantially lower than that of the ED for
the same $N$.

Fig.~\ref{fig:Pfa_all}-(b) illustrates $P_{\text{FA}}$ as a function
of the number of samples $N$ at a fixed SNR.
The ED false alarm probability converges to $0.5$ as $N$ increases.
This is due to that, with $N$ increase to infinity, the test statistic can be approximated by a normal distribution according to Central Limit Theorem. The mean and the variance of the ED are: 
	\begin{align}
		\nonumber
			\mu_0 &= 2\sigma^2 \\
			\nonumber
			\sigma_0^2 &= \frac{4\sigma^4}{N}
		\end{align} 
	separately. The threshold is also the noise energy $2\sigma^2$. Thus the false alarm rate when $N$ is vary large can be calculated by 
	\begin{align}
		\nonumber
			P_{\text{FA}}& = Q\left(\frac{2\sigma^2 - \mu_0}{\sigma_0} \right) = 0.5
	\end{align}
where \( Q(\cdot) \) is the Q function. This can be verified in Fig.~\ref{fig:Pfa_N}, the ED is converged to $\log\left({0.5}\right)$.

In contrast, the AD benefits from larger $N$:
its $P_{\text{FA}}^{(A)}$ decreases with $N$ because averaging more
amplitude samples sharpens the Rayleigh sum distribution and improves
separation between noise only and signal plus noise cases.

The RID curve lies between those of the ED and AD in both subplots,
in agreement with the sandwich relation
$P_{\text{FA}}^{(A)} \le P_{\text{FA}}^{(\text{RID})}
\le P_{\text{FA}}^{(E)}$ derived in Section~IV.
Near the ED/AD crossover region of interest, the RID false alarm
rate tracks that of the AD very closely, indicating that the
reliability based combining incurs only a mild false alarm penalty
relative to the high SNR detector.

\subsection{Misdetection Probability}
\begin{figure}[htbp]
	\centering
	\subfloat[$P_{\text{MD}}$ vs SNR ($N=16$).]{
		\includegraphics[width=0.95\linewidth]{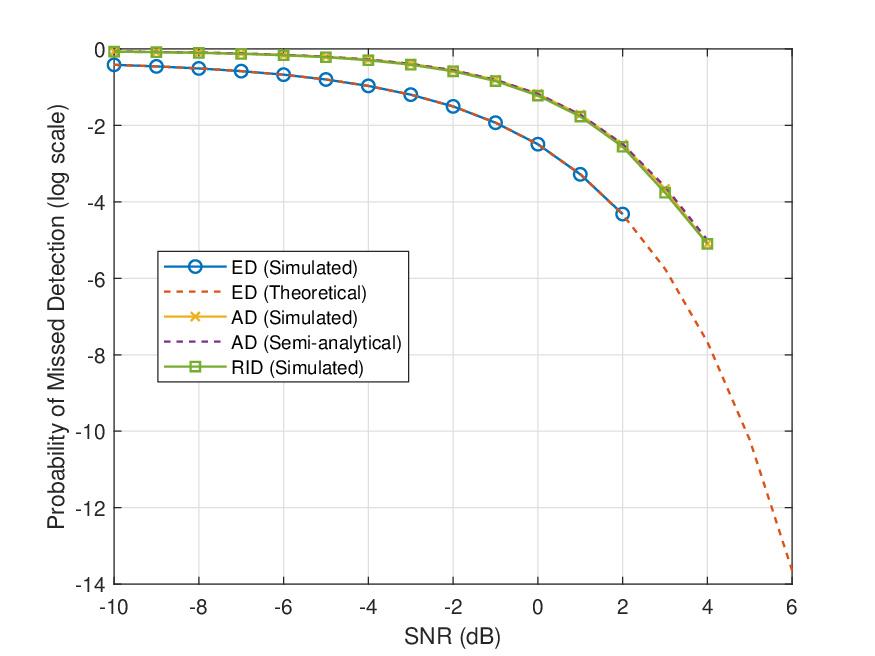}
		\label{fig:Pmd_snr}
	}
	
	\vspace{0.2cm} 
	\par
	
	\subfloat[$P_{\text{MD}}$ vs $N$ (SNR $=0$ dB).]{
		\includegraphics[width=0.95\linewidth]{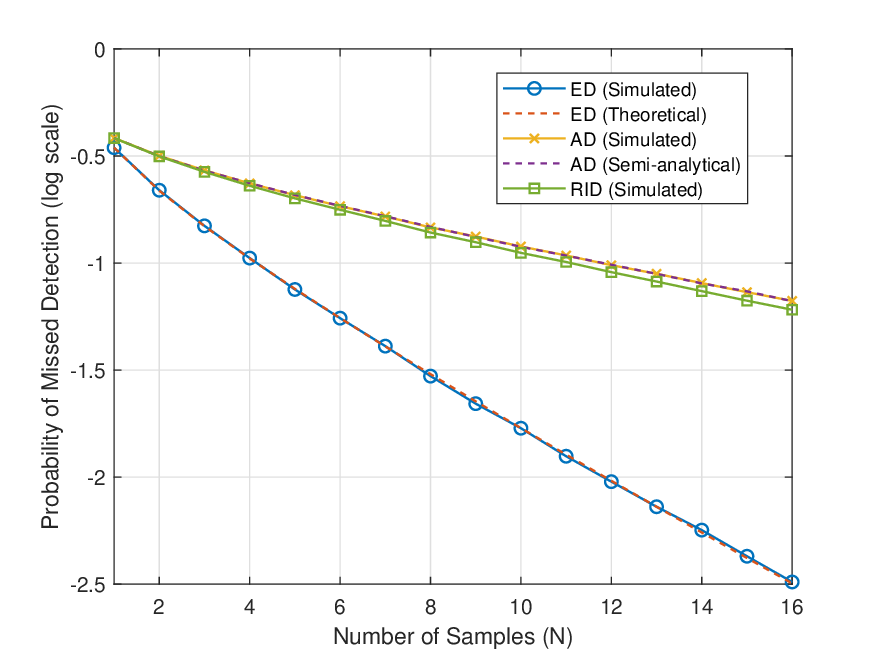}
		\label{fig:Pmd_N}
	}
	
	\caption{Simulated and analytical misdetection probability of ED, AD, and RID.}
	\label{fig:Pmd_all}
\end{figure}

Fig.~\ref{fig:Pmd_all}-(a) plots the $P_{\text{MD}}$ versus SNR for $N=16$. The simulated curves agree
well with the theoretical results,
confirming the accuracy of the noncentral chi-square and Rician-sum
based characterization. At low SNR, the ED attains a noticeably lower
$P_{\text{MD}}$ than the AD, whereas at high SNR the AD becomes
dominant. 

Fig.~\ref{fig:Pmd_all}-(b) shows $P_{\text{MD}}$ as a function of $N$
at $\text{SNR}=0$~dB. All detectors benefit from increasing $N$, but
the ED maintains a lower misdetection probability than the AD and the
gap widens with $N$. 

The RID curve lies slightly below that of the AD:
under $H_1$, disagreements are mainly of the form
Case (3), and in part of this region
the RID rule flips the AD decision to $H_1$ whenever the
estimated amplitude exceeds the switching threshold
$A_\star^{(3)}(N)$. These corrections reduce the AD misses without
altering decisions elsewhere, which explains why the RID curve remains
very close to, but consistently below, the AD curve.

\subsection{Probability of error and RID Gain}
\begin{figure}[htbp]
	\centering
		\centering
		\includegraphics[width=0.95\linewidth]{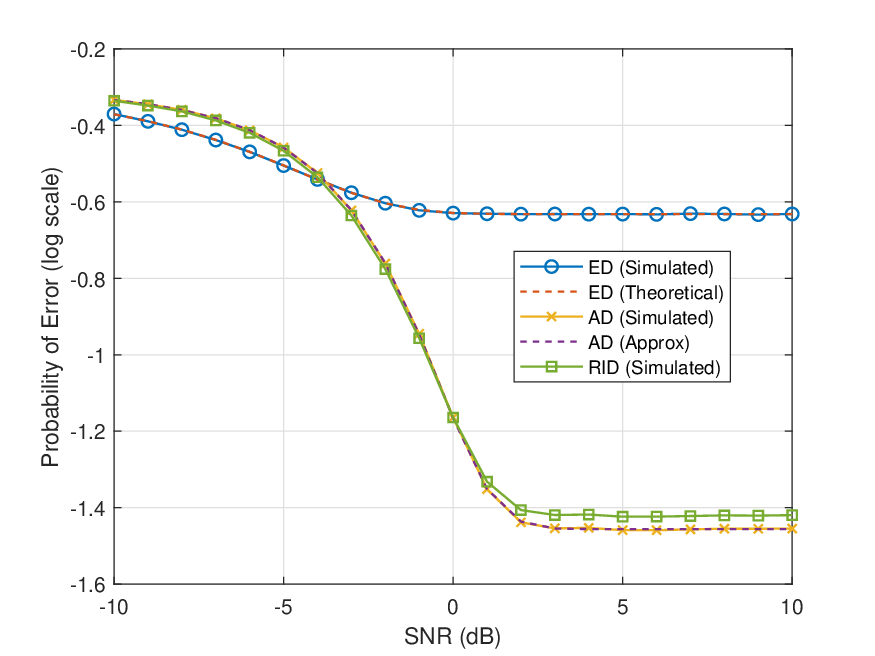}
		\label{fig:Pe_snr_16}
	\caption{Simulated and analytical total error probability of ED, AD, and RID versus SNR ($N=16$).}
	\label{fig:Pe_snr}
\end{figure}
\begin{figure}[htbp]
	\centering
	\subfloat[$P_e$ vs $N$ (SNR $=-5$ dB).]{
		\includegraphics[width=0.95\linewidth]{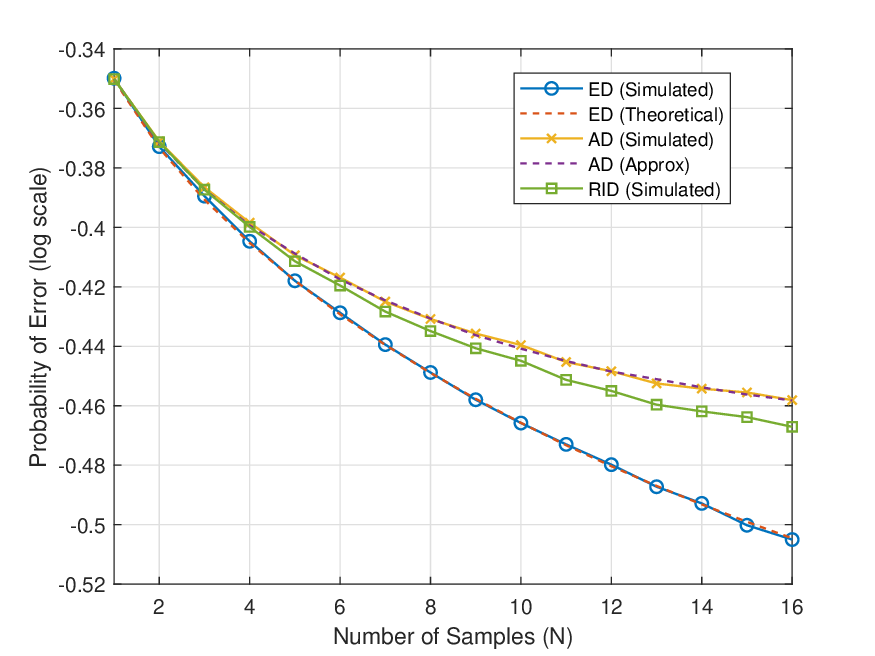}
		\label{fig:Pe_N_-5}
	}
	
	\vspace{0.2cm} 
	\par
	
	\subfloat[$P_e$ vs $N$ (SNR $=2$ dB).]{
		\includegraphics[width=0.95\linewidth]{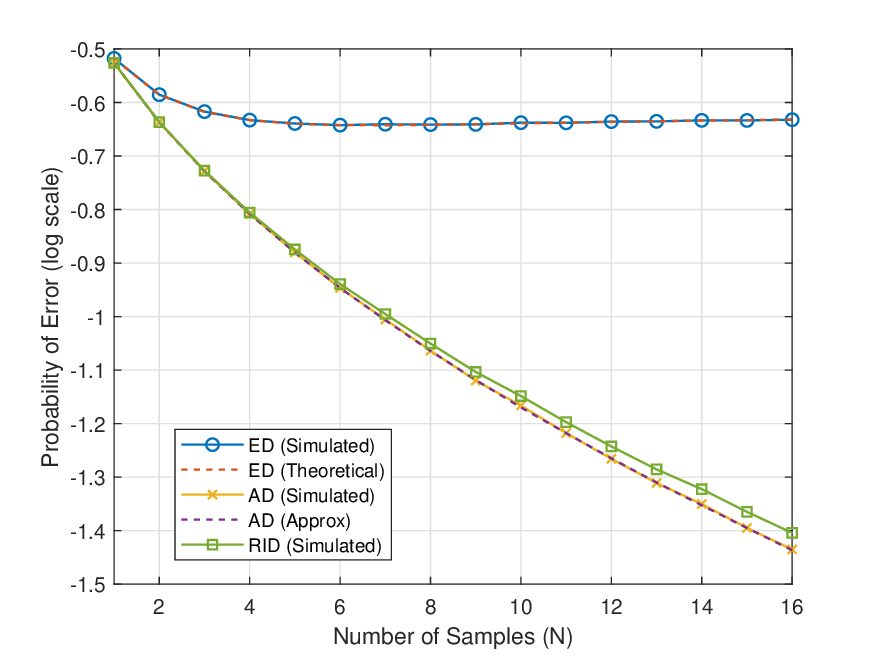}
		\label{fig:Pe_N_2}
	}
	
	\caption{Simulated and analytical total error probability of ED, AD, and RID versus $N$.}
	\label{fig:Pe_N}
\end{figure}

%
%
%
The overall tradeoff is captured by the total error probability
$P_e = \tfrac12(P_{\text{FA}}+P_{\text{MD}})$.
Fig.~\ref{fig:Pe_snr} plots $P_e$ versus SNR for $N=16$,
while Fig.~\ref{fig:Pe_N} shows $P_e$ versus $N$ at two representative 
SNR values (e.g., $\text{SNR}=-5$~dB and $2$~dB).

In Fig.~\ref{fig:Pe_snr}, the simulated $P_e$ curves of ED and AD are
almost indistinguishable from the analytical predictions, further
confirming the accuracy of the derived $P_{\text{FA}}$ and
$P_{\text{MD}}$ expressions. At low SNR, the ED exhibits a noticeably
lower error probability than the AD for all $N$, in line with its
low-SNR optimality. As SNR increases, the AD error probability drops
more steeply and eventually crosses that of the ED. This crossover SNR
moves to the left as $N$ increases, because both amplitude estimation
and reliability-based thresholding become more accurate with more
samples. At very high SNR, $P_{\text{MD}}$ becomes negligible and
$P_e$ approaches approximately $\tfrac12 P_{\text{FA}}$ for each
detector.

In Fig.~\ref{fig:Pe_N}, the dependence on the observation length is made
explicit. At $\text{SNR}=-5$~dB, the ED clearly dominates: its error
probability decreases monotonically with $N$, while the AD curve remains
higher despite also benefiting from a larger sample size. At
$\text{SNR}=2$~dB, the situation is reversed. The AD enjoys a
significant advantage and its $P_e$ decreases rapidly with $N$ as the
amplitude estimate becomes more reliable, whereas the ED is limited by
its false alarm floor: as $N$ grows, $P_{\text{MD}}^{(E)}$ shrinks,
but $P_{\text{FA}}^{(E)}$ converges to $0.5$, so $P_e^{(E)}$ tends
to $\tfrac12 P_{\text{FA}}^{(E)} \approx 0.25$.

Across both figures, the RID curves exhibit the robust behaviour
anticipated in Section~\ref{rid}. At low SNR, RID is outperforms AD slightly, since most
disagreements correspond to ED detecting while AD misses. As the SNR
increases, the RID performance gradually aligns with that of AD and
remains clearly better than ED, so that over most of the operating
range its error probability is close to the better of the two single
detectors rather than providing a large gain in any specific SNR
regime.

Overall, the numerical results confirm three key messages:
i) the ED is naturally the preferred detector in the low-SNR regime;
ii) the AD becomes advantageous in moderate-to-high SNR regimes and its
performance improves markedly with larger $N$; and
iii) the proposed RID rule provides a smooth and robust transition
between these regimes.

\section{Conclusion}
In this paper, we developed a unified framework for optimal noncoherent detection of constant-envelope signals from the magnitudes of the received samples. By leveraging on the simplified BFED structure, we sidestepped the complexities of traditional IQ-based receivers and phase/frequency synchronization. Through a rigorous Bayesian and GLRT-based analysis, we demonstrated that the optimal detector transitions between two regimes depending on the SNR for the minimum probability of error. In low SNR scenarios, the optimal solution reduces to an energy detector, providing a theoretical foundation for what had previously been a widely used heuristic for the first time. Under high SNR conditions, the optimum detector turns out to be a novel amplitude detector that uses the MLE of the amplitude to reliably distinguish signals from noise, further simplifying the receiver design.

Building on these results, we proposed a RID rule that adaptively fuses the ED and AD decisions using their analytically characterized error behaviors as reliability indicators. Numerical results confirm the accuracy and robustness of both detectors and show that the RID rule achieves consistent gains over either detector alone and a smooth performance transition over the entire SNR range. The closed-form approximations and derived distributions also enable efficient performance evaluation.

Overall, the proposed framework not only offers a theoretically optimal and practically implementable detection paradigm but also provides new insights into fundamental relationships between energy detection and amplitude estimation. Further research may explore extensions to the optimal detection in an unknown noise scenario, or to the design of low complexity receivers for the mobile edge applications such as WSNs, small IoT devices and so on.

\bibliographystyle{IEEEtran}
\bibliography{reference_abbreivated.bib}

\end{document}